\documentclass[preprint,trackchanges]{aastex631}
\usepackage{amsmath}
\usepackage{xcolor}

\newcommand{\al}[1]{\textcolor{teal}{#1}}

\begin{document}

\title{Metal-Enriched Atmospheres in Warm (Super- and Sub-)Neptunes Induced by Extreme Atmospheric Escape}

\correspondingauthor{Amy Louca}
\email{louca@strw.leidenuniv.nl}

\author[0000-0002-3191-2200]{Amy J. Louca}
\affil{SRON Netherlands Institute for Space Research, Niels Bohrweg 4, 2333 CA Leiden, The Netherlands}

\author{Yamila Miguel}
\affiliation{SRON Netherlands Institute for Space Research, Niels Bohrweg 4, 2333 CA Leiden, The Netherlands}
\affiliation{Leiden Observatory, Leiden University, 
Einsteinweg 55, 2333 CC Leiden
The Netherlands}






\begin{abstract}

Planet formation impacts exoplanet atmospheres by accreting metals in solid form, leading to atmospheric C/O and S/N ratios that deviate from their host stars.
Recent observations indicate differing metal abundances in planetary atmospheres compared to their stellar companions (e.g., \citeauthor{Alderson2022} \citeyear{Alderson2022}; \citeauthor{Bean2023} \citeyear{Bean2023}). However, these observations are biased toward mature planets, raising questions about whether these abundances result from formation or evolved over time. Another way to alter an atmosphere is through the qescape of particles due to thermal heating. This study examines how billions of years of particle escape affect metal abundances. 
Using an adjusted stellar evolution code incorporating hydrodynamic escape, we model a warm ($T_{\mathrm{eq}} \approx 1000$ K) super-Neptune-type planet ($M_{\mathrm{ini}} = 26$ $M_{\oplus}$) orbiting a solar-type star. Our results show increased metal-to-hydrogen abundances of $\sim$50-70x initial enrichment after 10 Gyr. We also see a 0.88x decrease in C/O abundance and a 1.27x increase in S/N abundance, which can affect the interpretation of planet formation parameters.
We also simulate the evolving atmosphere using chemical kinetics and radiative transfer codes, finding substantial increases in SO$_2$, CO$_2$, and H$_2$O abundances and a decrease in CH$_4$ abundance. These changes are easily observable in the IR waveband transmission spectrum. Our findings demonstrate that extreme escape of lighter particles significantly influences the evolution of warm Neptunes and complicates the interpretation of their observational data. This highlights the need to consider long-term atmospheric evolution in understanding exoplanet compositions.

\end{abstract}



\section{Introduction} \label{sec:intro}

It is believed that the formation sites of exoplanets leave an imprint on their composition (\citeauthor{Oberg2011} \citeyear{Oberg2011}). 
The temperature gradient within protoplanetary disks creates ice lines at various distances from the host star, leading to the accretion of heavy atoms in solid, molecular form during planet formation. Consequently, in planetary atmospheres, this changes metal (i.e., atoms heavier than H and He) ratios, such as the carbon-to-oxygen ratio (C/O). Observations have shown that the metal abundance of planetary atmospheres is indeed different with respect to the composition of their host star (e.g., \citeauthor{Alderson2022} \citeyear{Alderson2022}, \citeauthor{Bean2023} \citeyear{Bean2023}). However, a substantial part of these planets have evolved for several billions of years, raising the question of whether this observed heavy metal abundance is an imprint from formation or has changed over time. 

There are several ways to potentially alter a planetary atmosphere over time, one of which is the escape of particles due to thermal heating. Exoplanets with short orbital periods endure extreme escape throughout their lifetime. XUV heating from their host star results in an upward bulk motion of particles within the atmospheres, known as \textit{hydrodynamic escape} and is the main driver of extreme atmospheric evaporation (\citeauthor{Johnstone2015} \citeyear{Johnstone2015}). With sufficient energy from the stellar XUV irradiation, the velocity of the particles exceeds the escape velocity, $v_{\mathrm{esc}}$, and eventually surpasses the \textit{Roche-lobe radius}. At this point, the particles become unbound to the planet. Light particles, such as hydrogen and helium, suffer most from this escape mechanism. In some cases, the escape of these light gases is so extreme that they can drag along heavy metals (i.e, particles with an atomic mass larger than helium). This was first observed by \citet{VidalMadjar2004} for the escape of oxygen and carbon on HD 209458 b and subsequently also for other metals such as Si, S, and Fe on various types of hot gaseous planets (see e.g., \citeauthor{Linsky_2010} \citeyear{Linsky_2010}; \citeauthor{Schlawin2010} \citeyear{Schlawin2010}; 
 \citeauthor{DosSantos2023} \citeyear{DosSantos2023}). The amount of metals escaping depends on the momentum gained by the drag of the extreme hydrogen upflow. To obtain enough momentum for exceeding $v_{\mathrm{esc}}$, the downward gravitational force should be smaller than the upward heating force. 
This gravitational force is directly proportional to the escaping body mass, $F_g \propto m$ (where $F_g$ is the gravitational force and $m$ is the escaping body mass). Higher particle masses, therefore, need more heat transfer to obtain the same escape velocity as lower particle masses. Assuming that the heat transfer has the same efficiency for all particles, heavier particles are expected to have a less extreme escape. Naturally, a discrepancy emerges between the relative particle abundances, and we expect to see fractionation between metals of different masses. 

This has been theoretically worked out by \citet{Hunten1987}, where they defined the \textit{crossover mass} to include the drag of heavy elements in their mathematical framework of particle escape. In this work, they focused on hydrogen-dominated atmospheres of Mars/Earth-like size to explain the currently found mass-dependent depletion of the noble gases within the atmospheres of Mars and Earth. Their results showed that a substantial mass fractionation of heavy gases is expected to emerge due to hydrodynamic escape. This could result in changing relative metallic content within exoplanet atmospheres. 
In contrast, \citet{Mordasini2016} showed that formation is the dominating process in defining planetary spectra. They included both formation and evolution models of the planet that take envelope evaporation into account. Their main focus is, however, on relatively massive hot Jupiters (see also \citeauthor{Louca2023} \citeyear{Louca2023}), and it is argued in their study that envelope evaporation is not the dominating process. They, therefore, do not include mass fractionation on these planets and use formation models to explain their planetary spectra. 

Working further on these results, we investigate the effect of hydrodynamic escape on the mass fractionation of metals like carbon (C), nitrogen (N), oxygen (O), sulfur (S), magnesium (Mg), silicon (Si), and iron (Fe) within a warm super-Neptune orbiting a solar-like star that could be easily observed using current state-of-the-art observing facilities. Using planet evolution models that take extreme atmospheric escape into account, we look at the effect of metal drag on the mass fractionation of heavy metals to ultimately find out whether formation creates an imprint on spectral features or whether we should consider the atmospheres to be more dynamic over time than initially thought. This is important in explaining and correctly interpreting currently observed planetary spectra. 
In the next section, we lay out the methodology and then discuss the results in sections 3 and 4.

\section{Methodology} \label{sec:methods}

The methodology in this study has two main steps. First, planetary evolution is simulated to evaluate how extrinsic parameters such as mass, radius, and metallicity evolve (section \ref{sec:planetary_evo}). These parameters are then forwarded to a radiative transfer and chemical kinetics code to see how the atmosphere evolves consequently (section \ref{sec:spectral_evo}).

\subsection{Planetary evolution}
\label{sec:planetary_evo}

\subsubsection{Metal drag}
The planetary interior and envelope are evolved by making use of an adapted version of the \textit{Modules for Experiments in Stellar Astrophysics}, \texttt{MESA} (\citeauthor{mesa1} \citeyear{mesa1}; \citeyear{mesa2}; \citeyear{mesa3}; \citeyear{mesa4}; \citeyear{mesa5}; \citeauthor{mesa6} \citeyear{mesa6}). Hydrodynamic escape is included in the form of a \textit{hydro-based approximation}, HBA (\citeauthor{kubyshkina2018} \citeyear{kubyshkina2018}; \citeyear{kubyshkina2020}). We make use of the MIST database (\citeauthor{Dotter2016} \citeyear{Dotter2016}; \citeauthor{Choi2016} \citeyear{Choi2016}; \citeauthor{mesa1} \citeyear{mesa1}; \citeyear{mesa2}; \citeyear{mesa3}) to update the stellar flux as a function of time within the \texttt{MESA} simulations. We obtain time-dependent planet characteristics such as its mass and radius from these simulations. 

Using these characteristics, the metallicity of the envelope is evaluated posteriorly (see \citeauthor{Louca2023} \citeyear{Louca2023}). 
We assume a simple model of metal drag to estimate the mass fractionation of metals. The crossover mass from the model dictates whether metal drag must be factored in. If the mass of the minor species within the atmosphere is lower than the crossover mass, it will be dragged along with the bulk motion of the major escaping species. Contrariwise, if the mass of the secondary species is higher than the crossover mass, it will remain in the atmosphere. 
The crossover mass is defined as

\begin{equation}
    m_c = m_1 + \frac{kTF_1}{bgX_1}
    \label{eq:mc}
\end{equation}
where $m_c$ is the crossover mass, $m_1$ is the mass of the primary species (i.e., hydrogen), $T$ is the equilibrium temperature, $F_1$ is the escape flux of the primary species, $X_1$ is the mixing ratio of the bulk gas in the atmosphere, $k$ is the Boltzmann constant, $b$ is the binary diffusion parameter, and $g$ is the gravitational acceleration of the planet.

The escape flux of the primary species is dependent on the received stellar XUV flux, which relaxes over time (see the detailed formulation in \citeauthor{kubyshkina2018} \citeyear{kubyshkina2018}, hereafter K18). This causes the crossover mass to decrease as the system evolves and slowly becomes smaller than the atomic mass of the secondary species (see Fig. \ref{fig:mc_evo}). We, therefore, expect to see less metal drag over time, depending on the atomic mass of the secondary species. Heavier particles, such as sulfur, will fall above the crossover mass threshold sooner than lighter particles, such as nitrogen, oxygen, and carbon. This discrepancy in the degree of metal drag will, therefore, lead to a relative change in abundance between metals. 

We simulate a warm super-Neptune, with an initial mass of $M = 26$ $M_{\oplus}$ and an initial envelope mass fraction of $f = 0.55$. We assume a solid rocky core that consists of silicates and heavier metals, similar to Earth's core. The envelope is assumed to be hydrogen-dominated, with solar abundance, and homogeneously mixed.
The planet orbits a solar-like host star at 0.075 au, giving it an equilibrium temperature of $T_{\mathrm{eq}} \sim 1000$ K.

Finally, metal drag highly depends on equilibrium temperature. 
This is because the escape flux of the major species is, in the least extreme scenario, proportional to $\dot{M_1} \propto T_{\mathrm{eq}}^{-3.7489}$ (see K18), and the escape flux of the secondary species is directly proportional to the escape flux of the primary species. Hence, changing the temperature alters metal drag by almost four orders in magnitude.

Furthermore, a decrease in temperature also lowers the crossover mass, which results in lower escape ages. To investigate the influence of temperature on the metal enhancement within atmospheres, we vary the semi-major axis between 0.075 - 0.15 au, giving a temperature range of $T_{\mathrm{eq}} \approx$ 700-1000 K and a binary diffusion parameter range of $b/10^{19} \approx$ 3.22 - 3.98 cm$^{-1}$ s$^{-1}$ (following \citeauthor{Marrero2009} \citeyear{Marrero2009}).

\subsubsection{Fractionation: time-dependent escape flux}
\label{sec:fractionation}
Metal fractionation is calculated by considering an upward metal flux that changes over time and depends on the major escaping species in the atmosphere. In particular, this upward metal flux depends on the upward flux of the major escaping species and their abundance (see chapter 5 of \citeauthor{Catling2017} \citeyear{Catling2017}),

\begin{equation}
    F_2 = \frac{n_2}{n_1}F_1\left(\frac{m_c-m_2}{m_c-m_1}\right)
    \label{eq:escape_flux}
\end{equation}

where subscripts 1 and 2 denote the major and minor species respectively. In this case, the major species that suffers from hydrodynamic escape is hydrogen, and the minor species that get dragged along with it are the metals carbon, nitrogen, oxygen, and sulfur. Here $F$ is the escape flux, \al{$n$} is the number density, and $m$ is the species mass. Note that the volume factor of the number densities cancels out in equation \ref{eq:escape_flux}, and we can, thus, also consider $N$ to be the full content of particles within the envelope of planets. Similarly, the flux terms both depend on the same surface area of the planet, and we can consider those as the mass loss in grams per second
\begin{equation}
    \dot{M}_2 = \frac{N_2}{N_1} \dot{M}_1 \left(\frac{m_c - m_2}{m_c - m_1}\right)
\end{equation}

The mass loss of the main escaping species due to hydrodynamic escape processes, $\dot{M}_1$, is calculated using the \textit{hydrodynamic escape approximation} from K18,
\begin{equation}
    \dot{M}_1 = e^{\beta} (F_{\mathrm{XUV}})^{\alpha_1} \left(\frac{a}{\mathrm{au}}\right)^{\alpha_2}\left(\frac{R_{\mathrm{pl}}}{R_{\oplus}}\right)^{\alpha_3}\Lambda^{K}
\end{equation}

where the parameters $\beta$, $\alpha_1$, $\alpha_2$, and $\alpha_3$ can be found in table 1 of K18, $F_{\mathrm{XUV}}$ is the XUV irradiation flux from the host star, $a$ is the orbital distance, $R_{\mathrm{pl}}$ is the planetary radius, $\Lambda$ is the Jeans escape parameter, and $K$ describes how fast the mass-loss rates decrease with increasing $\Lambda$, as defined in K18. This expression is an analytical approximation of the actual hydrogen mass loss based on a grid of hydrodynamical simulations.
The fractionation between metals can be calculated using this upward escape flux of metals due to drag. In particular, the number of metal particles is updated using
\begin{equation}
    N_{z_{t}} = N_{z_{t-1}} - \frac{1}{m_z} \int_{t-1}^{t} \dot{M}_2(t) \mathrm{d}t 
\end{equation}
where $m_z$ is the mass of the metal. The metal mass fraction is then updated by summing over all updated total metal masses 

\begin{equation}
    Z_t = \frac{\sum_{z} N_{z_t} \cdot m_{z}}{M_{\mathrm{env}}}
\end{equation}
where $M_{\mathrm{env}}$ is the envelope mass of the planet. The mass fraction of hydrogen is then updated by assuming that the hydrogen-to-helium mass fraction (X/Y) remains the same, which is a valid assumption for Neptune-sized planets that have a large envelope fraction (\citeauthor{Malsky2020} \citeyear{Malsky2020}). Since the hydrogen-to-helium fraction remains the same and all mass fractions must add up to 1 ($X + Y + Z = 1$), we update the hydrogen mass fraction using
\begin{equation}
    X_t = \frac{1-Z_t}{1+1/C}
\end{equation}
where $C$ is the initial hydrogen-to-helium fraction. For these calculations to hold, we use the condition that hydrogen always needs to be the most abundant species in the atmosphere, i.e.,
\begin{equation}
    X_t > Z_t
    \label{eq:condition}
\end{equation}
If this condition is no longer satisfied, we will regard the atmospheric escape as too extreme, and the planet will not be considered within the grid simulation.

Finally, the hydrogen content is updated by
\begin{equation}
    N_{H_t} = \frac{X_t\cdot M_{\mathrm{env}}}{m_H}
\end{equation}

 For initial conditions, we assume solar-like abundance with $X_{t=0}$ = 0.74 and number density ratios from \cite{Asplund2009}. We only consider the most abundant volatile and metal species, C, N, O, Mg, Si, S, and Fe.

 \subsection{Atmospheric evolution}
\label{sec:spectral_evo}
We continue this modeling setup for a planet with initial parameters of M = 26 M$_{\oplus}$  and $f_{\mathrm{env}}$ = 0.55 that orbits a solar-type star at 0.075 au. All planet characteristics, such as mass, radius, and metallicity, are updated according to the results from the planetary evolution code. 

\subsubsection{Temperature-Pressure profiles}

The thermal structures in the planet's atmosphere are updated using the open source radiative transfer code \texttt{HELIOS} (\citeauthor{Malik2017} \citeyear{Malik2017}; \citeyear{Malik2019}). The thermal profile of the planet is evaluated in six steps, during which the planet's radius, mass, and metallicity are updated according to all previous results. We keep stellar parameters constant to evaluate only the impact of hydrodynamic escape on the atmosphere. All included opacity sources are listed in table \ref{tab:linelists}. We consider pressure broadening due to H$_2$ collisions for the atoms Na and K up until 100 bar. The k-tables are sampled for a wavelength range of 0.06 - 200 $\mu$m, with a resolution element of $\lambda/\triangle \lambda = 1000$, and assuming chemical equilibrium for all species abundances using the open source chemical equilibrium code \texttt{FastChem} (\citeauthor{Stock2018} \citeyear{Stock2018}). Furthermore, we assume an internal temperature of $T_{\mathrm{int}} = 200$ K, a global heat redistribution factor of 0.33, and isotropic irradiation from the host star.

\subsubsection{Chemical evolution}

After having obtained the temperature profiles of the planet, the composition of the atmosphere is calculated using the (photo-)chemical kinetics code VULCAN (\citeauthor{Tsai17} \citeyear{Tsai17}; \citeyear{Tsai21}). The chemical composition of the atmosphere is also determined for six timesteps. The planet's radius, mass, metallicity, and thermal profile are updated at each time step. 
Metals included in the chemical network are C, N, O, and S. The chemical network comprises 574 forward- and backward reactions and 68 photochemical reactions. For the photochemical reactions, we include the stellar spectrum of the sun from \cite{Gueymard2003}. Like before, all abundances are initialized with the open-source chemical equilibrium code \texttt{FastChem}. Vertical mixing is considered in the form of both molecular- and eddy diffusion. For the eddy diffusion constant, we use a constant profile of $K_{zz} = 10^{10}$, which has been previously used by other studies (e.g., \citeauthor{Moses2013b} \citeyear{Moses2013b}; \citeauthor{Parmentier2013} \citeyear{Parmentier2013}; \citeauthor{Miguel2014} \citeyear{Miguel2014}).

\subsubsection{Transmission spectra}

For the final modeling step, the different atmospheres are forwarded to the radiative transfer code petitRADTRANS (\citeauthor{prt2019} \citeyear{prt2019}; \citeyear{prt2020}; \citeauthor{prt2022} \citeyear{prt2022}) to obtain the transmission spectra at each of the six timesteps. All spectra are created assuming the opacity of the species H$_2$O, CH$_4$, SO$_2$, CO$_2$, CO, NH$_3$, H$_2$S, and H$_2$ (as listed in table \ref{tab:linelists}), and include collisional induced absorption (CIA) of H$_2$-H$_2$ and H$_2$-He. The resolution element for all opacity sources is $\lambda / \triangle \lambda = 1000$. 
 
\section{Results} \label{sec:results}

This section presents the findings related to planetary evolution and its impact on metal enhancement (section \ref{sec:plan_evo}). Following this, we focus on atmospheric evolution and its observational implications (section \ref{sec:atmos_evo}).

\subsection{Planetary evolution}
\label{sec:plan_evo}

\begin{figure}
    \centering
    \includegraphics[width=0.48\textwidth]{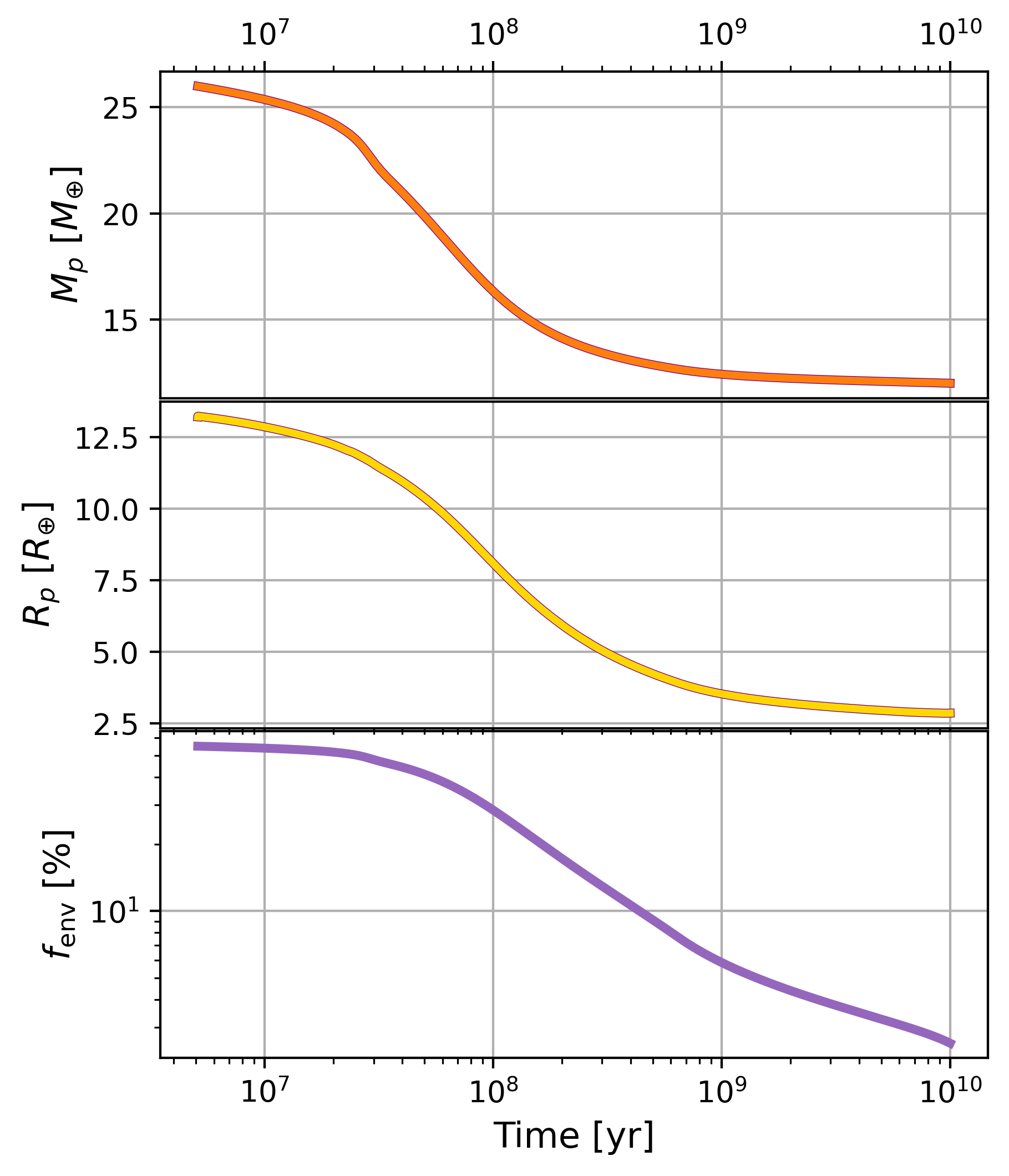}
    \caption{The evolution of the mass (upper panel), radius (middle panel), and envelope mass fraction (lower panel) of a planet with an initial mass of $M = 26$ $M_{\oplus}$ and initial envelope mass fraction of $f_{\mathrm{env}} = 55$\% orbiting a solar-like host star at a distance of $a = 0.075$ au.}
    \label{fig:MR_evo}
\end{figure}

\subsubsection{Mass, radius, and envelope}

The planet in this study is born a super-Neptune. Due to its relatively high equilibrium temperature ($\sim$1000K) and high initial envelope mass fraction, the planet suffers from great mass losses of its hydrogen-dominated envelope, as seen in figure \ref{fig:MR_evo}. After $\sim$1-2 Gyr, the super-Neptune has evolved into a sub-Neptune-type planet. Most of its mass is lost in the first $\sim$20-100 Myr of the evolution, which is also the period it contracts most. During this period, the extreme escape of hydrogen is expected to drag along most metals. After 10 Gyr, the final density has converged to $\sim$2.85 g/cm$^3$, with an envelope mass fraction of $\sim$2.5\%, leaving the planet with a planed, but still observable, gaseous envelope.

\subsubsection{Metal mass loss}
\label{sec:massloss}

\begin{figure}
    \centering
    \includegraphics[scale=0.5]{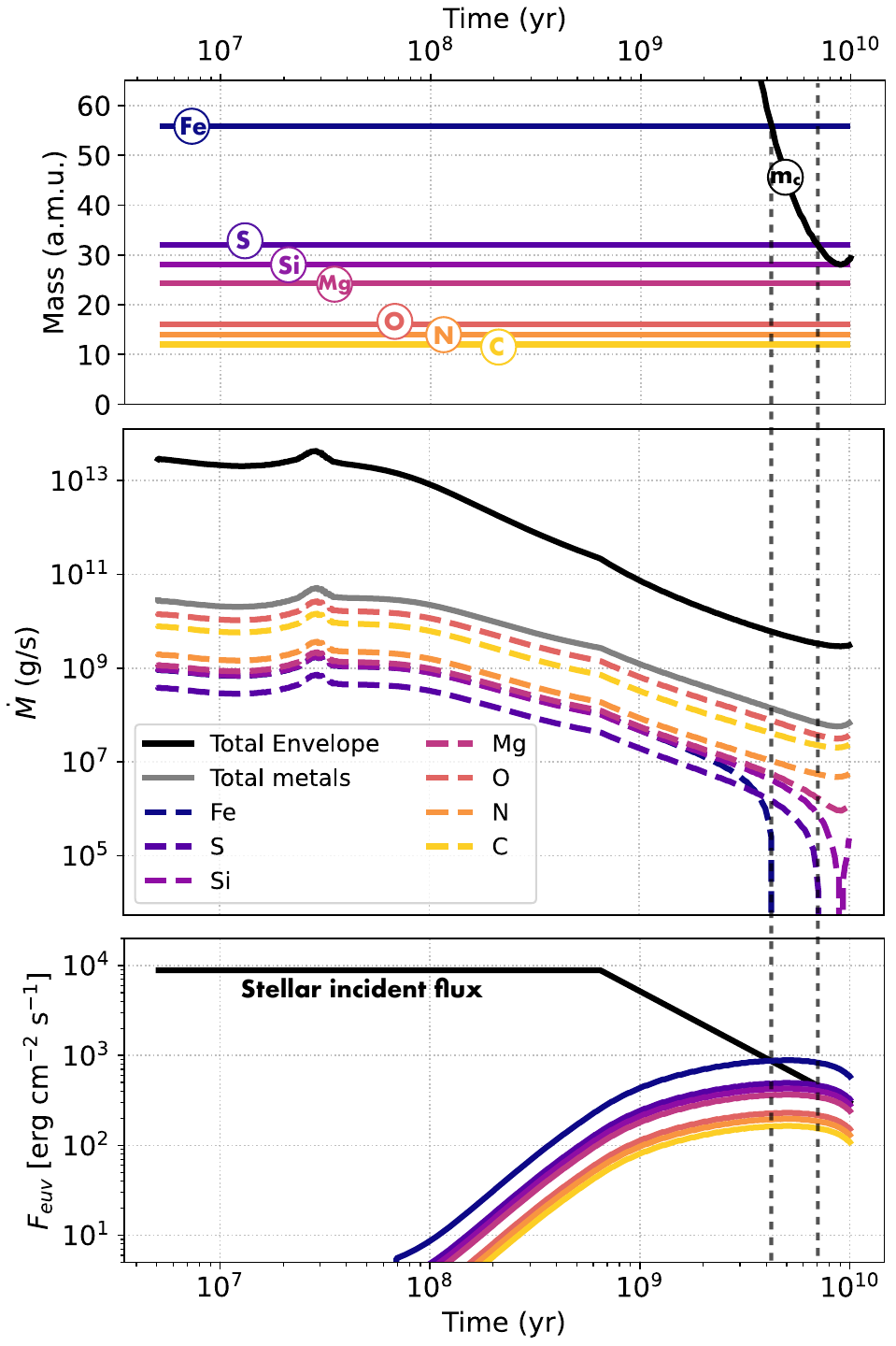}
    \caption{Evolution of the crossover mass (black solid line, top-panel), the evolution of the total mass loss of each element (middle panel), and the evolution of the incoming $F_{euv}$ irradiation of the host star (bottom panel) for a planet with mass $M = 26 \mathrm{\ M_{\oplus}}$, an initial envelope fraction $f_{\mathrm{env}} = 0.55$, and a semi-major axis of $a = 0.075$ au. The horizontal colored lines represent the different metallic species considered in this study (i.e., Iron, Sulfur, Silicon, Magnesium, Oxygen, Nitrogen, and Carbon) as indicated by colors. The vertical dashed lines indicate the escape age of Fe and S (see text for more details). In the bottom panel, the colored lines represent the minimum stellar flux needed for these species to be dragged with the escaping hydrogen.}
    \label{fig:mc_evo}
\end{figure}

The mass loss of each metallic species due to drag depends on 3 factors, one of which is the same for each metal (i.e., the escape flux of the minor species, see eq. \ref{eq:escape_flux}). However, each metal's atomic mass and relative abundance differ, which causes a difference in final mass loss per atom. 
The middle panel of figure \ref{fig:mc_evo} illustrates this difference in mass loss for each of the species. The relative difference between the total envelope mass loss and the metal mass loss (grey line) decreases over time, from $\sim$3 orders in magnitude to $\sim$2 orders in magnitude. 

Both the \textbf{sulfur} and \textbf{iron} mass loss show an asymptotic behavior around $\sim$7-8 Gyr and $\sim$4-5 Gyr, respectively, wherein they drop to zero. This is because, 
 around these periods, the crossover mass falls below the atomic mass of sulfur and iron, as seen in the top panel of figure \ref{fig:mc_evo}. At the same time, it can be seen in the bottom panel that the minimum stellar flux needed to drag these elements is higher than the incident stellar flux at that time. This indicates that hydrogen does not get enough stellar energy to drag along elements such as iron and sulfur.
 Similarly, the mass loss of \textbf{silicon} also shows a steep drop around $\sim$ 9 Gyr but continues to escape shortly after. This is due to the slightly increasing crossover mass at $\sim$ 9 Gyr. Lighter metals such as \textbf{oxygen}, \textbf{nitrogen}, and \textbf{carbon} do not show this asymptotic trend since they always fall below the crossover mass during the evolution and will, therefore, always be dragged along and escape. This contrasting behavior between metals is thought to cause a relative enhancement in heavier metals, such as sulfur and iron.

The grey line shows the total metal mass loss in figure \ref{fig:mc_evo}.
Notably, \textbf{oxygen} contributes significantly to the overall metal loss, surpassing the mass loss of nitrogen and carbon, despite its higher atomic mass. This is because the initial abundance of oxygen is relatively high compared to all other metals considered, which shows that initial abundance is a key factor in determining the mass loss evolution of metals.

\newpage
\subsubsection{The metal-to-hydrogen enhancement over time}
\label{sec:metal_to_hydrogen}

\begin{figure}
    \centering
    \includegraphics[scale=0.5]{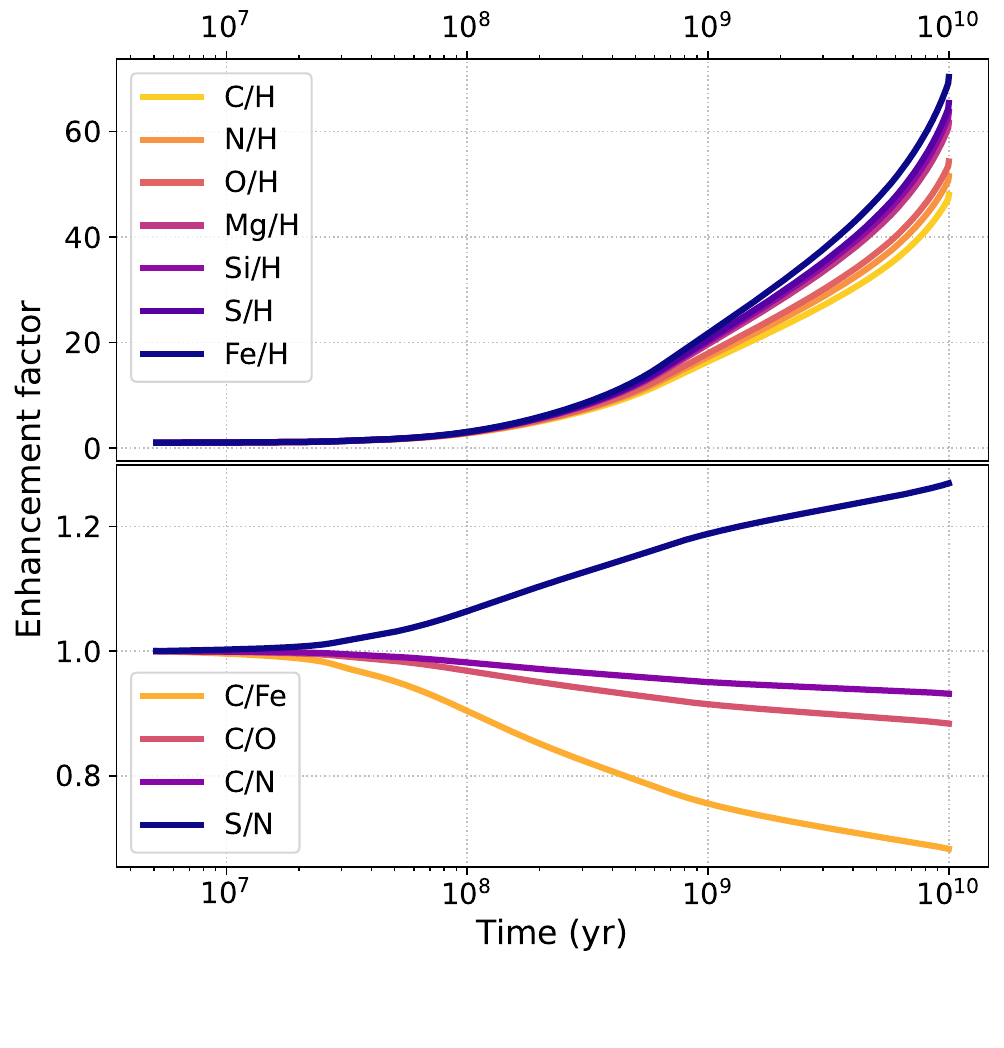}
    \caption{The temporal evolution of the metal-to-hydrogen enhancement (top panel), and of the metal-to-metal enhancement (bottom panel). Note that not all metal-to-metal ratios are shown here, but only the most relevant cases are depicted.} 
    \label{fig:fractionation}
\end{figure}

\begin{figure}
    \centering
    \includegraphics[scale=0.45]{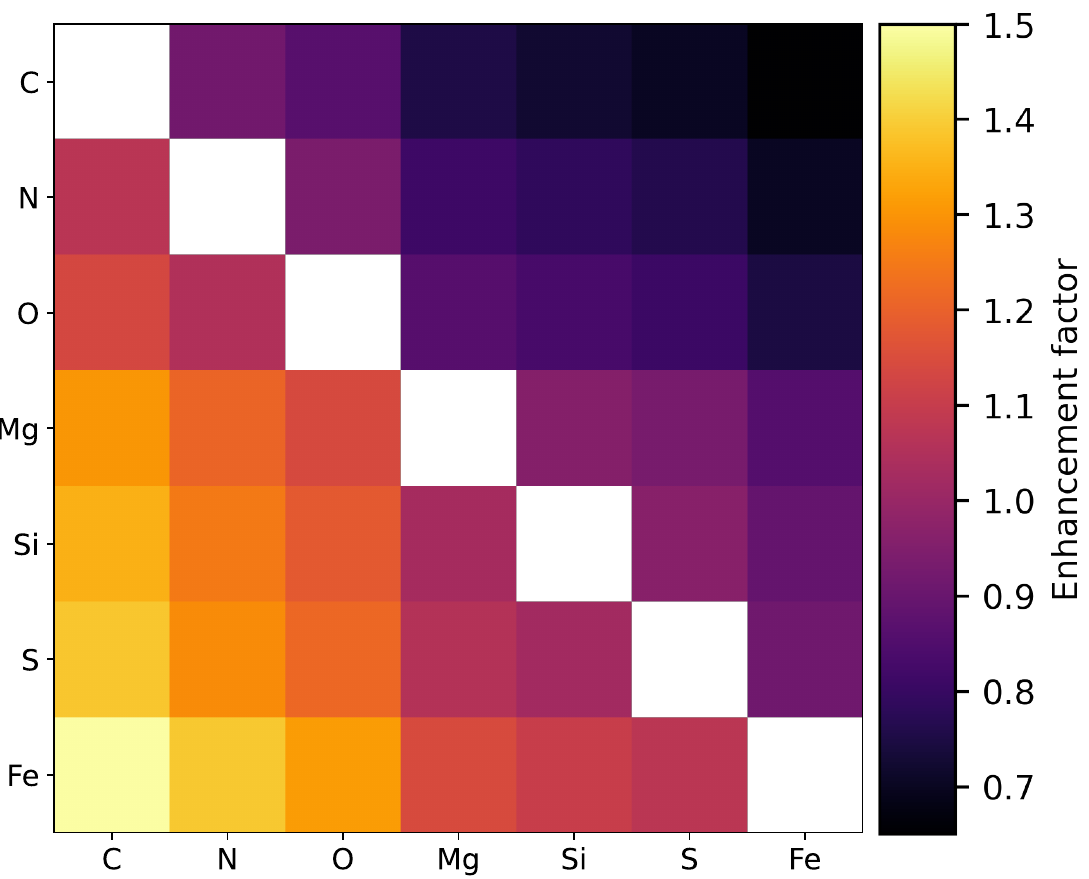}
    \caption{The metal-to-metal enhancement due to metal drag for all the metals considered in this study after 10 Gyr into the simulation. The horizontal and vertical axes denote the metals included and are sorted by their atomic mass.}
    \label{fig:metal-metal-ratios}
\end{figure}

Considering all previous results, we anticipate observing a substantial increase in the metal-to-hydrogen ratios. Illustrated in the top panel of figure \ref{fig:fractionation} is the temporal relative metal enhancement. In this case, metal enhancement means the evolving metal-to-hydrogen ratio normalized by its initial value.
This figure shows that most of the metal enhancement should occur after 0.1 Gyr into the simulation, hinting that the first 100 million years are a window into observing the imprints of planet formation.  Shortly after, the metal enhancement increases drastically for all species considered in this study. With a final enhancement of $\sim$48x the initial abundance, the least affected species is \textbf{carbon}. The most affected species, \textbf{iron}, shows an increase factor of $\sim$70x its initial abundance. This indicates that there is a correlation between atomic metal mass and metal enhancement. The heavier metals (e.g., Si, S, and Fe) have a higher final enhancement over time than the lighter metals (e.g., C, N, and O). 

\subsubsection{The metal-to-metal enhancement over time}
\label{sec:metal_to_metal}

Due to each metal's different mass loss rate and escape age, the metal-to-metal ratio is also expected to change over time. The bottom panel of figure \ref{fig:fractionation} illustrates the temporal metal-to-metal ratio for the most extreme and relevant metal ratios. Contrarily to before, it can be seen that metal fractions have already started to change in the first 100 Myrs of evolution. This is due to each species' different mass loss and initial abundance (see section \ref{sec:massloss}). After this period, the metal-to-metal enhancement seems to converge for each species. The maximum enhancement can be found in the \textbf{C/Fe} ratio, with a 0.68x decrease after 10 Gyr.  

The \textbf{C/O} ratio is widely used to indicate where the planet formed and is often thought to be static in time. These results show, however, that we expect to see a decrease of 0.88x after 10 Gyr. Depending on the initial C/O, this decrease in ratio can have implications on inferring the formation sites of planets (\citeauthor{Oberg2011} \citeyear{Oberg2011}). A value of C/O$\geq 1$ implies that the planet has formed around $a\geq10$ au. A decrease factor of 0.88x due to evolution can thus temper the inference of formation parameters and should be accounted for. 
Also considering the possibility of non-conventional C/O ratios resulting from in-situ formation near the soot line (\citeauthor{Bergin2023} \citeyear{Bergin2023}), the challenge of linking C/O ratios to planetary formation and evolution is further complicated by the impact of extreme atmospheric escape.

Another indicator for formation sites of exoplanets is the \textbf{S/N} ratio. In particular, it has been stated that exoplanets are thought to have formed far away from the host star when S/N $>$ C/N (e.g., \citeauthor{Turrini2021} \citeyear{Turrini2021}; \citeauthor{Pacetti2022} \citeyear{Pacetti2022}). In this study, we see, however, that planetary evolution can result in an elevated S/N abundance ($\sim1.27$x initial abundance) and a depletion in the C/N abundance ($\sim0.93$x initial abundance), perhaps leading to falsely believe that the planet has endured long-term migration. 

The final enhancements of all metal combinations are illustrated in figure \ref{fig:metal-metal-ratios}. This figure shows a color gradient as a function of atomic metal mass: the higher the atomic metal mass ratio discrepancy, the higher the enhancement factor becomes. However, this assumes that the planet's envelope and atmosphere are well-mixed and no heavy metal particles sink to the inner parts of the planet due to their relatively high atomic mass.

 \begin{figure}
    \centering 
    \includegraphics[width=0.5\textwidth]{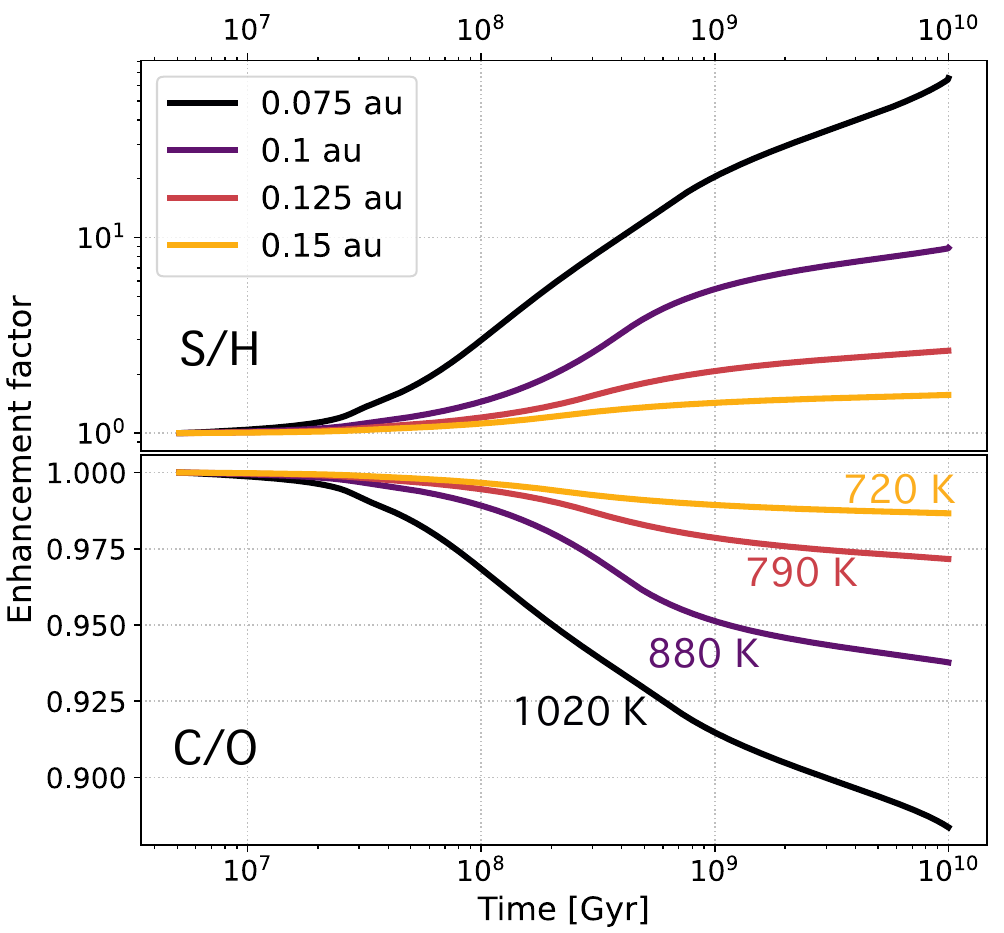}
    \caption{The evolution of the enhancement factors of the sulfur-to-hydrogen ratio (upper panel) and carbon-to-oxygen ratio (lower panel) for four different orbital separations, ranging between 0.075 au and 0.15 au.}
    \label{fig:enhancement_oneplanet}
\end{figure}

\subsubsection{The impact of orbital distance}
\label{sec:all_planets}
 The effect of the temperature - and, therefore, semi-major axis - on metal enhancement is shown in figure \ref{fig:enhancement_oneplanet} (top-panel). This figure shows the sulfur-to-hydrogen enhancement as an example case. The enhancement factor decreases substantially when increasing the semi-major axis. This is due to the decreasing irradiation flux from the host star and, therefore, the decreasing equilibrium temperature, limiting the hydrodynamic escape of hydrogen. Even though this would also suggest less metal drag, the metal enhancement is still limited by the restricted amount of hydrogen escape.
 Similarly, lower equilibrium temperatures also suppress the metal-to-metal ratios in change. The lower panel of figure \ref{fig:enhancement_oneplanet} shows the evolution of the \textbf{C/O} ratio for four different semi-major axes. For higher semi-major axes and, thus, lower equilibrium temperatures, the change in the C/O abundance becomes less extreme. 

\begin{figure}
    \centering
\includegraphics[width=0.48\textwidth]{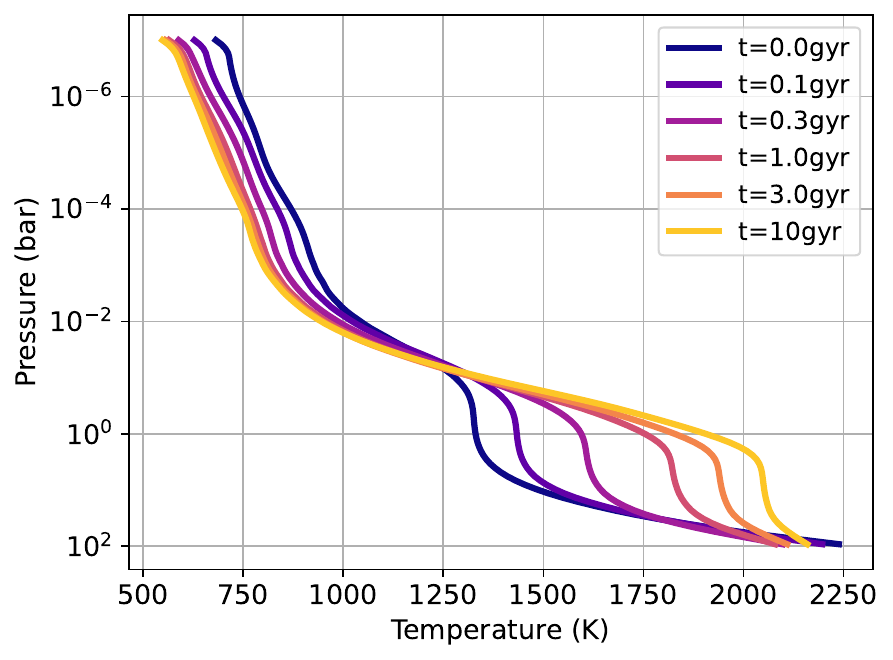}
    \caption{Evolution of the TP-profile for several timesteps, as indicated by colors.}
    \label{fig:tp_evo}
\end{figure}

\subsection{Atmospheric evolution}

\label{sec:atmos_evo}

\subsubsection{Thermal evolution}

The full thermal evolution of the atmosphere is shown in figure \ref{fig:tp_evo}. The biggest atmospheric changes are found in the upper ($P < 10^{-2}$ bar) and lower ($P > 10^{-2}$ bar) regions. Particularly, the upper region shows a decrease in temperature over time, whereas the lower region shows a substantial increase in temperature over time. The upper region shows a change of $\sim$250 K in 10 billion years, whereas the lower region shows a change of $\sim$750 K. 

The increasing metallicity due to extreme atmospheric escape can explain the rising temperature in the lower atmosphere. This causes the atmosphere to contain more \textbf{H$_2$O} and \textbf{CO$_2$}, which are greenhouse gases that heat up the atmosphere over time. In particular, this greenhouse effect makes the atmosphere more opaque in the deeper regions around $P \approx$ 0.1 - 1 bar.

To conserve the global energy within the system, the upper atmosphere cools in response to this increasing temperature in the deeper regions. 

\begin{figure*}
    \centering
    \includegraphics[width=1.0\textwidth]{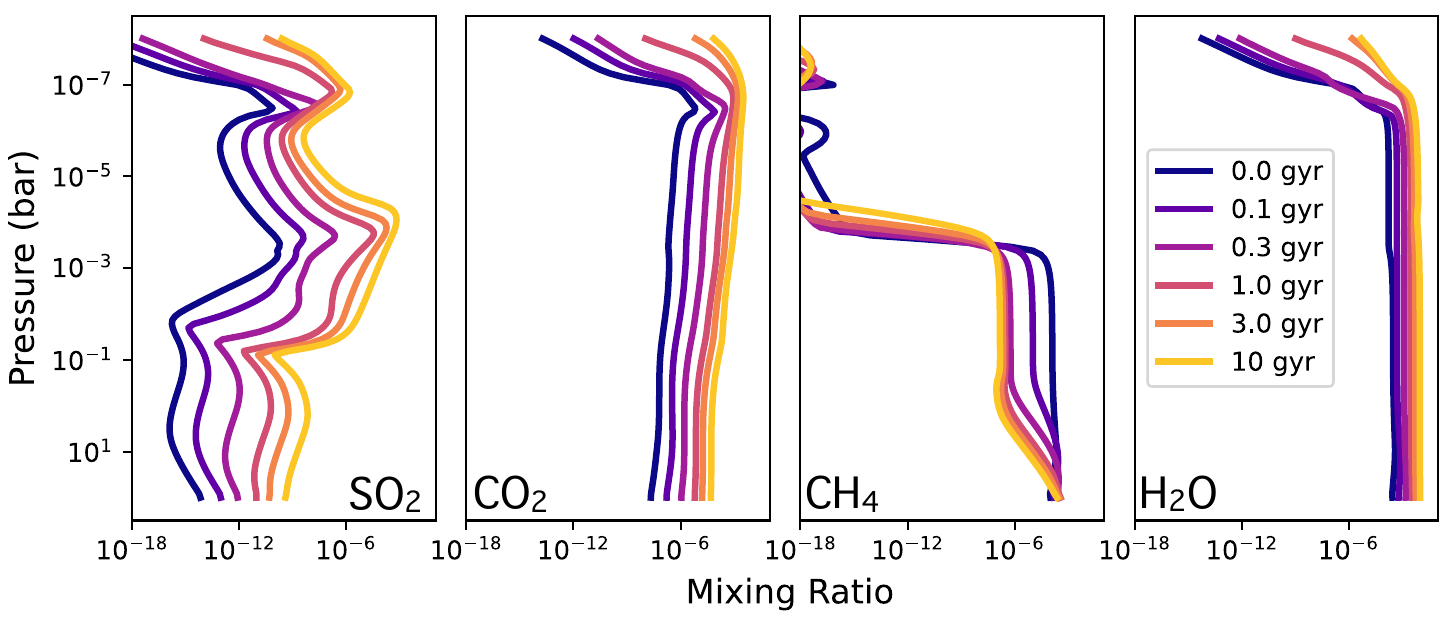}
    \caption{The evolution of the mixing ratio of SO$_2$, CO$_2$, CH$_4$, and H$_2$O for a planet with an initial mass of $M = 26$ $M_{\oplus}$ and an initial envelope fraction of $f_{\mathrm{env}} = 0.55$. The colors in each subplot illustrate the timesteps.}
    \label{fig:chemical_evo}
\end{figure*}

\subsubsection{Chemical evolution}

The previously presented changes in metallicity, metal-to-metal ratios, and temperature can lead to compositional changes within planetary atmospheres over time. The full chemical evolution of the species SO$_2$, CO$_2$, CH$_4$, and H$_2$O is illustrated in figure \ref{fig:chemical_evo}. Most changes occur between 0.1 and 1.0 Gyr in the evolution of all species. This is when the metallicity changes from $\sim$3xSolar to $\sim$22xSolar. The biggest change can be found in the \textbf{SO$_2$} abundance.  This relative species abundance increases $\sim$ 7 orders in magnitude over 10 billion years in nearly all regions of the atmosphere. Both the increase in total metallicity and the increase in the sulfur-to-metal enhancement contribute to this drastic enrichment of SO$_2$ in the atmosphere. This change in abundance is greatest in the upper atmosphere, with an increase of 11 orders in magnitude.  Using a photochemical kinetics model, \cite{Polman2023} showed that high metallicity atmospheres are expected to contain a significant amount of SO$_2$ abundance, and this was, shortly after, also shown observationally by \cite{Tsai2023}. Another indicator of high metallicity atmospheres is the high \textbf{CO$_2$} abundance (e.g., \citeauthor{Lodders2002} \citeyear{Lodders2002}; \citeauthor{Zahnle2009} \citeyear{Zahnle2009}; \citeauthor{Moses2013b} \citeyear{Moses2013b}). From figure \ref{fig:chemical_evo}, we can see an increase of 3 orders in magnitude in the relative CO$_2$ abundance throughout the atmosphere due to the increasing metallicity. In the upper atmosphere, this abundance increase goes up to 9 orders in magnitude. The relative \textbf{H$_2$O} abundance shows the least change throughout the atmosphere, with an increase of $\sim 1$ - 2 orders in magnitude, which can still be regarded as a significant increase in relative abundance. The biggest changes in H$_2$O can be found in the upper atmosphere, where we see an increase of almost 9 orders in magnitude. 
Finally, in contrast to the other species, \textbf{CH$_4$} shows a depletion of $\sim4$ orders in magnitude in the lower atmosphere. For the deeper regions ($P > 10^{-1}$ bar), this is caused by the drastic increase in temperature in the lower atmosphere, as seen in figure \ref{fig:tp_evo}, in combination with vertical mixing and the fact that an increase in metallicity implies less hydrogen available to form methane. 

\subsubsection{Spectral evolution}

\begin{figure*}
    \centering
    \includegraphics[width=1.0\textwidth]{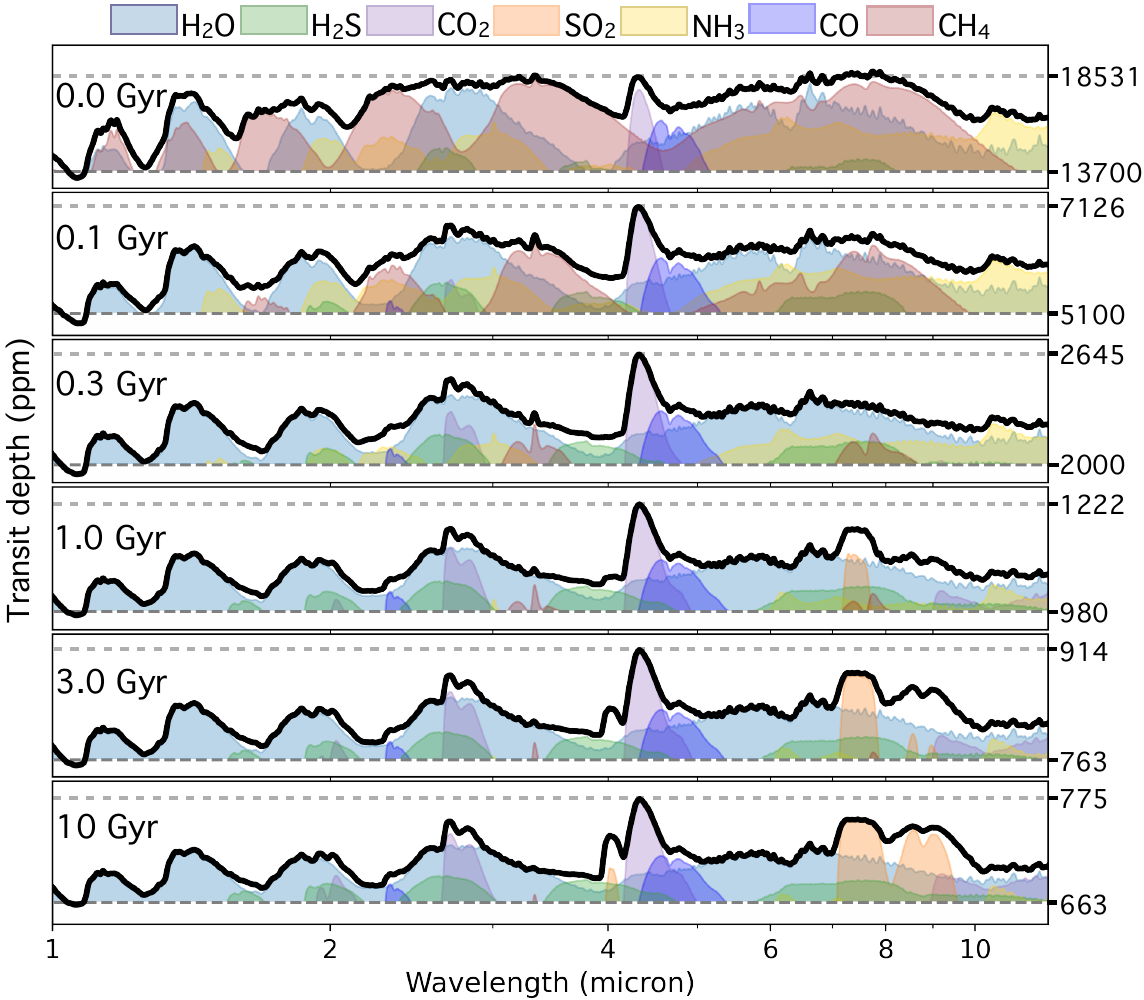}
    \caption{The evolution of transmission spectrum evaluated at 0, 0.1, 0.3, 1, 3, and 10 Gyr, including the contribution of each species as indicated by color.}
    \label{fig:transmission_evo}
\end{figure*}

The evolution of the transmission spectrum is illustrated in figure \ref{fig:transmission_evo}. The biggest changes in spectra can be seen between 0.1 and 1.0 Gyr. During this period, methane-dominated features are transitioned to a mixture of metallic features such as SO$_2$, CO, and H$_2$S. 
Additionally, the water features become more prominent due to increased water abundance over time. More explicitly, we see over time that

\begin{itemize}
    \item \textbf{0.0 - 0.1 Gyr (M/H $\sim$ Solar):} \textbf{CH$_4$} is the dominating species and most prominent in the spectrum. There are other hints of \textbf{H$_2$O} in the lower wavelength region ($<$ 3 $\mu$m), \textbf{CO$_2$} at $\sim$4.4$\mu$m, and \textbf{NH$_3$} in the long wavelength region ($>$10$\mu$m).
    \item \textbf{0.1 - 0.3 Gyr (M/H $\sim$ 3-8xSolar):} The low-metallicity features of CH$_4$ (2 - 8 $\mu$m) and NH$_3$ ($>10$ $\mu$m) start to disappear. Slowly, some higher metallic species such as \textbf{H$_2$S} (at $\sim$3.8$\mu$m) become more apparent.
    \item \textbf{0.3-1.0 Gyr (M/H $\sim$ 8-22xSolar):} All low-metallicity features, including NH$_3$, have disappeared from the spectrum. In the long wavelength region, the \textbf{SO$_2$} features at $\sim$7-8 $\mu$m start to appear. Additionally, a small \textbf{CO} feature appears around $\sim$ 5 $\mu$m.
    \item \textbf{1 - 3 Gyr (M/H $\sim$ 22-38xSolar):} The SO$_2$ features become significant within the spectrum at $\sim$4 $\mu$m and $7-8$ $\mu$m. 
    \item \textbf{3 - 10 Gyr (M/H $\sim$ 38-70xSolar):} There is little spectral change besides the bigger high-metallicity features. 
\end{itemize}

Overall, the most significant changes are found in the decreasing CH$_4$ features at 3-4 $\mu$m, which disappear after $\sim$0.1 Gyr after formation, and the increasing SO$_2$ features at 4 and 7-9 $\mu$m. \\

\begin{figure*}
    \centering
    \includegraphics[width=1.0\textwidth]{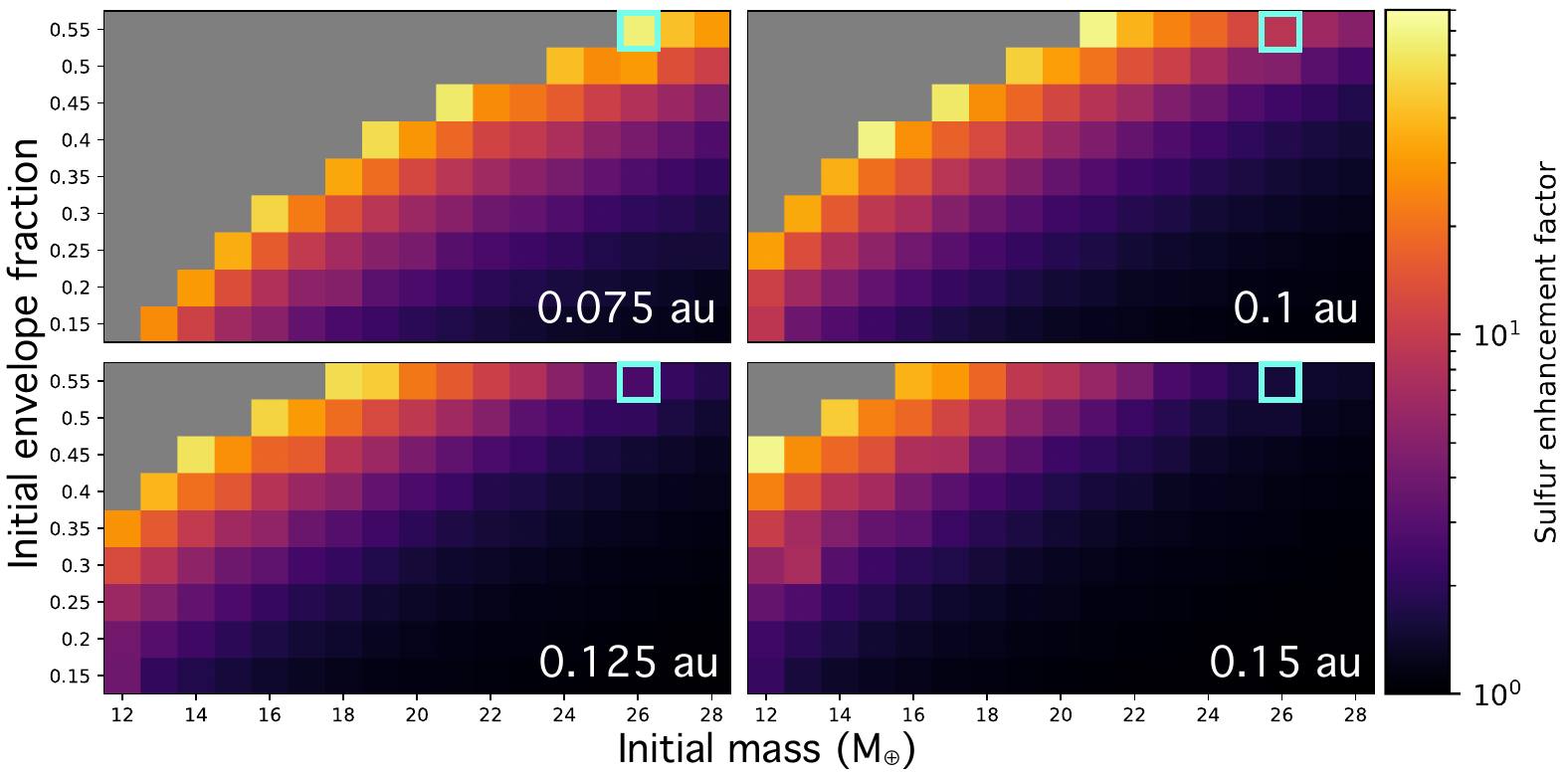}
    \caption{The sulfur enhancement (i.e., $\frac{(S/H)_t}{(S/H)_0}$) evaluated after 10 Gyr for various initial masses, envelope fractions, and orbital distances within the grid space. Each subplot represents the sulfur enhancement for a different orbital distance (0.075 - 0.15 au), as labeled within the figure. The grey spaces are simulations for which the hydrogen depletion is too strong, and equation \ref{eq:condition} no longer holds. The blue square in each subpanel indicates the previously studied planet with an initial mass of $M = 26$ $M_{\oplus}$ and an initial envelope fraction of $f = 0.55$.}
    \label{fig:S_H_evo}
\end{figure*}

\section{Discussion}
This research showed that the influence of extreme atmospheric escape can lead to metal enhancement and fractionation of species for a specific case of a warm super-Neptune. More specifically, the metal-to-hydrogen ratios increased by a factor up to $\sim$50-70x, depending on metallic species and planetary system. These results are, however, limited to a few assumptions.

Firstly, the planets were assumed to have a rocky inner core and a homogeneously mixed envelope, though it has been shown in other studies that a compositional gradient might fit better for the planets in our own solar system (e.g., \citeauthor{Wahl2017} \citeyear{Wahl2017}; \citeauthor{Mankovich2021}\citeyear{Mankovich2021}; \citeauthor{Miguel2022} \citeyear{Miguel2022}). With such planetary structures, extreme escape from the outer metal-poor atmosphere could expose the inner metal-rich atmosphere, where metal drag becomes less significant. 
Additionally, metal-enriched envelopes are expected to enhance planetary contraction due to the increased mean molecular weight. In this study, however, metal enrichment was calculated in a posterior manner and, therefore, did not give any feedback to the planetary evolution. 

We did not account for stellar evolution or changes in the planet's internal temperature in our atmospheric calculations, though these factors likely alter the TP-profile, chemistry, and transmission spectra. Additionally, evolving internal temperatures influence the dominant escape mechanism. \citet{Modirrousta-Galian2023} classifies extreme atmospheric escape into time-dependent regimes, with core-powered mass loss driven by residual internal heat. This process can lead to significant atmospheric loss, particularly for planets with thin atmospheres or close to their host star. \citet{Gupta2019} showed that the \textit{radius valley} can be explained by core-powered mass models. While our study focused on photoevaporation, atmospheric evolution may differ when core-powered mass loss is considered.

Finally, we limited this study to one extreme scenario: a born super-Neptune that evolves into a sub-Neptune over time due to hydrodynamic escape. To see whether the results of extreme metal enhancement due to planetary evolution are limited to only this planet, we performed a grid study where we varied the initial masses (12 - 26 $M_{\oplus}$) and envelope mass fractions (0.15 - 0.55). Figure \ref{fig:S_H_evo} illustrates the sulfur-to-hydrogen enhancement for the full grid of planets. Each subpanel here also indicates the super-Neptune as presented in sections \ref{sec:plan_evo} - \ref{sec:atmos_evo} with the blue square. Here, we can see that only a select number of planets experience metal enhancement due to hydrodynamic escape, depending on the equilibrium temperature. The super-Neptune of $M = 26$ $M_{\oplus}$ and $f = 0.55$ shifts outside of this region for lower equilibrium temperatures. Overall, we see that for larger semi-major axes, the relative sulfur enhancement becomes less substantial for planets with intermediate-to-high masses and low-to-intermediate envelope fractions. This is due to the lower temperatures, causing less extreme hydrogen escape and, therefore, less metal drag overall. 

\begin{figure*}
    \centering
    \includegraphics[scale=0.95]{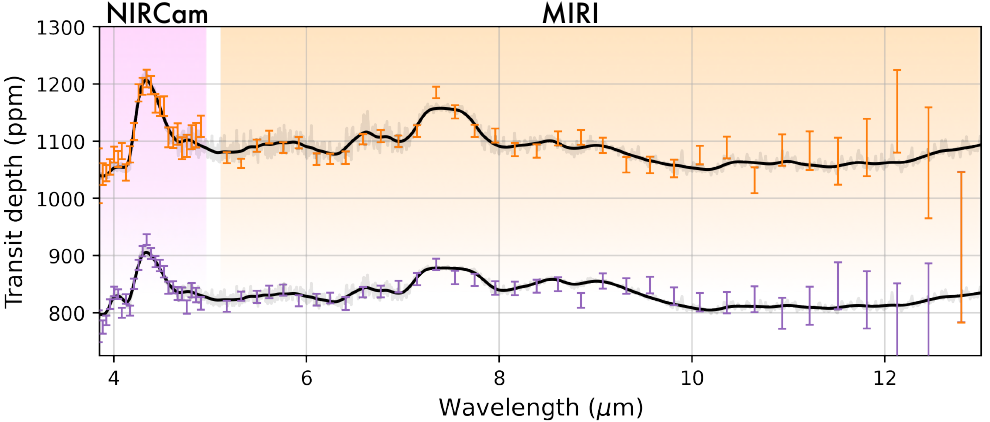}
    \caption{Synthetic (black lines) and synthetic JWST data (orange and purple) for NIRCam F444W and MIRI LRS. The upper transmission spectrum represents the young planet at $t=$ 1 Gyr (orange datapoints), and the lower transmission spectrum represents the matured planet at $t=$ 3 Gyr (purple datapoints).} 
    \label{fig:pandexo_JWST}
\end{figure*}

This study also showed that the drastic increase in relative metal abundance can be directly seen in observational transmission spectra in the IR waveband. Such an effect of extreme atmospheric escape on atmospheric composition can be tested observationally by comparing the IR transmission spectrum of adolescent planets with comparable mature planets. Finding suitable targets might still be challenging due to the lack of extremely young planets. This study showed, however, that these planet targets do not necessarily have to be extremely young. A substantial change can still be found in the transmission spectrum between 1 - and 3 Gyr. To asses whether these changes in spectra are observable with current state-of-the-art space missions, we create synthetic JWST data using \texttt{pandexo} (\citeauthor{Batalha2017} \citeyear{Batalha2017}). The most promising features to detect for metallicity enhancement are SO$_2$, CO$_2$, and CO between 4 - 10 micron (see figure \ref{fig:transmission_evo}). The instruments covering these features are the NIRCam F444W mode (4-5 $\mu$m) instrument and the MIRI Slitless LRS mode (4.9 - 27.9 $\mu$m) instrument. The predicted transmission spectra for these two instruments are created for a 1 Gyr and 3 Gyr planet (from figure \ref{fig:transmission_evo}), as shown in figure \ref{fig:pandexo_JWST}. For both spectra, the datapoints were created using 8 transits for a planet with a transit duration of 0.13375 days orbiting a solar-type star (J-magnitude of 7.7). The CO$_2$ and SO$_2$ features at $\sim$ 4.3 $\mu$m and $\sim$ 7.8 $\mu$m respectively are clearly visible for both spectra. However, the more pronounced SO$_2$ feature at $\sim$ 4 $\mu$m can only be detected for the more matured planet, as seen in the purple datapoints of figure \ref{fig:pandexo_JWST}. This feature can be used to track the evolution of exoplanets.

\begin{table*}[ht]
    \centering
    \begin{tabular}{c|c|c|c|c}
        \textbf{Molecule} & \textbf{Temperature range (K)} & \textbf{Pressure range (bar)} & \textbf{Line lists \texttt{HELIOS}} & \textbf{Line lists \texttt{pRT}}  \\
        \hline
        \hline
        H$_2$O & 50 - 2900 & $10^{-8}$ - $10^3$ & POKAZATEL$^1$ & POKAZATEL$^1$ \\
        CH$_4$ & 50 - 2900 & $10^{-8}$ - $10^3$ & YT34to10$^2$ & YT34to10$^2$ \\
        SO$_2$ & 50 - 2900 & $10^{-8}$ - $10^3$ & ExoAmes (v2)$^3$ & ExoAmes (v2)$^3$\\
        CO & 50 - 2900 & $10^{-8}$ - $10^3$ & Li2015$^4$ & HITEMP$^{11}$ \\
        H$_2$S & 50 - 2900 & $10^{-8}$ - $10^3$ & AYT2$^5$ & AYT2$^5$\\
        CO$_2$ & 50 - 2900 & $10^{-8}$ - $10^3$ & CDSD-4000$^6$ & UCL-4000$^{12}$\\
        PH$_3$ & 50 - 2900 & $10^{-8}$ - $10^3$ & SAlTY$^7$ & N/A \\
        NH$_3$ & 50 - 2900 & $10^{-8}$ - $10^3$ & N/A & CoYuTe$^{13}$ \\
        H$_2$ & 50 - 2900 & $10^{-8}$ - $10^3$ & RACPPK$^8$ & N/A \\
        \hline
        \hline
        \textbf{Atom} & & & & \\
        \hline
        \hline
        H & 2500 - 6100 & $10^{-8}$ & VALD$^9$ & N/A\\
        He & 2500 - 6100 & $10^{-8}$ & Kurucz$^{10}$ & N/A\\
        Na & 2500 - 6100 & $10^{-8}$ & Allard19$^{14}$ & N/A\\
        K & 2500 - 6100 & $10^{-8}$ & Allard16$^{15}$ & N/A\\
    \end{tabular}
    \caption{The line lists used for each opacity species included in radiative transfer. [1] \citet{Polyansky2018}; [2] \citet{Yurchenko2014}; \citet{Yurchenko2017} [3] \citet{Underwood2016}; [4] \citet{Li_2015}; \citet{Somogyi2021}; [5] \citet{Azzam2016}; \citet{CHUBB2018178}; [6] \citet{TASHKUN20111403}; [7] \citet{Sousa-Silva2014}; [8] \citet{Roueff2019}; [9] \citet{VALD};[10] \citet{Kurucz1995}; [11] \citet{ROTHMAN20102139}; [12] \citet{Yurchenko2020}; [13] \citet{Phillip2019}; [14] \citet{Allard2019}; [15] \citet{Allard2016}}
    \label{tab:linelists}
\end{table*}

\section{Conclusion} \label{sec:conclusion}

In this study, we used an extended version of a planetary evolution code that includes the hydro-based approximation to evolve a warm ($T_{\mathrm{eq}} \approx 1000$ K) super-Neptune orbiting solar-like star for $\sim$ 10 Gyr. We used posterior calculations that include metal drag to estimate the metal enhancement for C, N, O, Mg, Si, S, and Fe. To summarise, the results showed that

\begin{itemize}
    \item heavier species experience less metal drag and, therefore, more metal-enhancement
    \item oxygen contributes most to the total metal mass loss due to its high initial abundance in a solar-like environment
    \item the lighter species carbon, nitrogen, and oxygen are dragged along by the hydrodynamic escape of hydrogen throughout the full evolution for higher-density
    planets
    \item a maximum metal-enhancement of $\sim$48-70x the initial metallicity is expected, depending on metal mass 
    \item a maximum metal-to-metal enhancement factor of 1.5x initial enrichment can be found in the carbon-to-iron ratio
    \item the C/O ratio is decreased by a factor of 0.88x due to planetary evolution and the S/N ratio is increased by a factor of 1.27x, both of which impact the interpretation of planet formation parameters
\end{itemize}

We also simulated the atmosphere at several timesteps using (photo-)chemical kinetics and radiative transfer codes to evaluate chemical composition and spectra evolution. Here, we found that

\begin{itemize}
    \item the upper atmosphere ($P < 10^{-2}$ bar) cools over time, whereas the lower atmosphere ($P > 10$ bar) heats up due to the increasing greenhouse gasses (i.e., H$_2$O and CO$_2$) in the atmosphere 
    \item the SO$_2$, CO$_2$, and H$_2$O abundances show an increase in the upper atmosphere over time due to the increasing metallicity. For SO$_2$, this increase can even be seen throughout the full atmosphere.
    \item most (spectral) changes can be seen after 0.1 Gyr into the evolution. For $t < 1$ Gyr, the atmosphere consists mainly of methane signatures due to the low metallicity. After this period, a wide variety of high-metallicity features can be seen, such as SO$_2$, CO, and H$_2$S

\end{itemize}

Overall, this study showed that the evolution of the metal ratios of (volatile) metals is significant for warm (super-)Neptunes with a substantial initial envelope fraction and should be considered when evolving the planet and inferring planet parameters. In several cases, the formation is expected to imprint the metal-to-metal ratios that remain the same throughout the remainder of the planet's lifetime. Nonetheless, this study showed that exoplanets born as (sub- and super-)Neptunes can experience drastic changes throughout their lifetime, and evolution should be considered. 

\bibliography{main}{}
\bibliographystyle{aasjournal}



\end{document}